\documentclass[12pt]{article}

\renewcommand{\thefootnote}{\fnsymbol{footnote}}
\usepackage{epsf}

\def\aa{{c_\infty}}
\def\bb{{c_0}}
\def\unit{{\bf \Lambda}}

\def\alh{{\widehat \alpha_s}}
\def\al{\alpha_s}
\def\lf{\ell}
\def\nc{N_c}
\def\nf{n_f}
\def\nfh{{\widehat {\nf}}}
\def\nch{{\widehat {\nc}}}
\def\cf{{C_F}}
\def\cfh{\widehat{C_F}}
\def\dabc{{(d_{abc}/4)^{2}}}
\def\dabcf{{\left({d_{abc}\over 4}\right)^{2}}}
\def\dabch{{({\widehat{d_{abc}/4})^{2}}}}
\def\dabcfh{{\widehat{\dabcf}}}

\def\index{{\varpi}}
\def\u1{U(1)}
\def\sun{SU(\nc)}
\def\un{U(\nc)}
\def\rl{{{\mathcal R}_{n,\lf}}}
\def\col#1#2{{{\mathcal C}_{#1}^{#2}}}
\def\colh#1#2{{\widehat \col{#1}{#2}}}
\def\dol#1#2{{{\mathcal D}_{#1}^{#2}}}

\def\onegraph{ {\bf \bigcirc} }
\def\twograph{ {\bf \bigcirc \!\! =\!= \!\!  \bigcirc}}
\def\threegraph{ {\bf \bigcirc \!\! =\!= \!\! \bigcirc \!\! =\!= \!\! 
\bigcirc}}
\def\sup#1#2#3{#1^{#2}_{#3}}

\begin{document}
\begin{flushright}
SLAC-PUB-7603\\
SCIPP 97/21\\
July 1997
\end{flushright}
\vfill

\begin{center}
{\LARGE {Aspects of $\sun$ Gauge Theories in the Limit
of Small Number of Colors}
\footnote{\baselineskip=13pt Work partially supported by the Department
of Energy, contract DE--AC03--76SF00515, and INT-9319788.}}\\

\vspace{15mm}
{\bf Stanley J. Brodsky}\footnote{sjbth@slac.stanford.edu}\\
\vspace{5mm}
{\em Stanford Linear Accelerator Center \\
Stanford University, Stanford, California 94309}

\vspace{15mm}
{\bf Patrick Huet}\footnote{huet@scipp.ucsc.edu}\\
\vspace{5mm} {\em Santa Cruz Institute for Particle
Physics \\
University of California, Santa Cruz, California 95064} \vfill
\end{center}

\begin{abstract}
{\baselineskip=13pt \noindent We investigate properties of the color 
space of $\sun$ gauge theories in the limit of small number of colors 
($\nc\to0$) and large number of flavors.  More generally, we introduce 
a rescaling of $\al$ and $\nf$ which assigns a finite limit to colored 
quantities as $\nc\to0$, which reproduces their known large-$\nc$ 
limit, and which expresses them as an analytic function of $\nc^2$ for 
arbitrary value of $\nc$.  The vanishing-$\nc$ limit has an Abelian 
character and is also the small-$\nc$ limit of $[\u1]^{\nc-1}$.  This 
limit does not have an obvious quantum field theory interpretation; 
however, it provides practical consistency checks on QCD perturbative 
quantities by comparing them to their QED counterparts.  Our analysis 
also describes the two-dimensional topological structure involved in 
the interpretation of the small $\nc$-limit in color space.}
\end{abstract}


\vfill
\newpage

\renewcommand{\thefootnote}{\alph{footnote}}
\setcounter{footnote}{0}

\subsection*{Introduction}

't Hooft has shown the utility of studying gauge theories with
internal symmetries in the limit of large number of
colors~\cite{thooft}.  In this note, we explore the analytic
properties of formulae for non-Abelian theories taking the number of
colors $\nc$ to zero and the number of flavors $\nf$ to infinity.  Our
underlying motivation is to attempt to uncover patterns in the group
structure of $\sun$ which encompasses a nontrivial set of values of
$\nc$.  Little is known about nontrivial values of $\nc$ except in
the large $\nc$-limit, where it has been shown to have an
interpretation in terms of a theory of
membranes~\cite{hoppetal}.\footnote{There is a conjecture that this
connection extends to finite values of $\nc$ in some supersymmetric
matrix models~\cite{susskind}.} The value $\nc = 0$ is another limit
where we can expect simplifications to take place, hence, to provide
additional clues as to what the theory is at finite $\nc$.  The limit
$\nc\to0$ has already found some applications in condensed matter
systems, in the context of spin glass~\cite{spinglass} and
localization phenomena~\cite{localization}.  Negative values of $\nc$
have been studied in a group-theoretic framework~\cite{cvita}.
Furthermore, numerical work has shown that QCD${}_{1+1}$ has a
well-defined particle spectrum in the limit $\nc,\,\nf/\nc\to
\infty$~\cite{engelhardt}, a particular limit first introduced by
Veneziano in QCD${}_{4}$~\cite{veneziano}.  The present analysis is to
our knowledge, the first attempt to implement the limit $\nc\to0$ in
the context of gauge theories.  In this investigation, we are not
concerned with the meaning of the corresponding spacetime quantum
theory but rather with the structure of the space of color degrees of
freedom.  Our analysis is preliminary, yet we find it suggestive of
the existence of a deeper mathematical structure which covers a larger
set of values of $\nc$.  Specifically, the vanishing-$\nc$ limit
singles out a point-like structure reflecting a cluster of interlinked
punctures on a surface, in contrast with the two-dimensional objects
characteristic of the large-$\nc$ limit.

To give meaning to a physical quantity as $\nc$ becomes vanishingly 
small, we need to provide a proper rescaling to the parameters of the 
theory.  Clearly, we need the gauge coupling $\alh \sim \al/\nc$, in 
order to compensate for the decrease the number of degree of freedom 
in the theory.  When fermions are included, we also need to compensate 
accordingly, $\nfh \sim \nf\nc$.  These choices are natural since, as 
we soon show, they provide a finite value to all physical quantities 
and constitute the counterparts of those choices made in the 
large-$\nc$ limit: $\alh \sim \al\nc, \nfh \sim \nf/\nc$ which, in 
this case, compensate for the ever-increasing number of fermionic and 
bosonic degrees of freedom~\cite{thooft,veneziano}.  As we argue soon, 
it is constructive to conjecture that these rescalings are special 
cases of a unique, yet unknown, analytic rescaling in $\nc$.  Such a 
universal rescaling would plausibly be of the form $\alh \sim \al \, 
f(\nc)/\nc, \nfh \sim \nf\,\nc/ g(\nc)$ with the functions $f(\nc)$ 
and $g(\nc) \sim \nc^2$ as $\nc$ becomes arbitrarily large and 
$f(0)\sim g(0) \sim \, const.$, in order to satisfy the above 
requirements.

\subsection*{Examples}

We illustrate the above setting with a few examples of the behavior of
QCD perturbative physical quantities under the tentative
rescaling\footnote{The more general rescaling $\unit_s$, $\unit_f$
would not change any of our conclusions.}
\begin{equation}
\alh = \al \times\,{\unit\over 2 \nc} \hskip2cm 
\nfh =  \nf \times\, {\nc\over \unit} .\label{rescaling}
\end{equation} 
where $\unit$ is an unknown function of $\nc$ satisfying ($\aa$, $\bb$ are
nonzero integers)
\begin{equation}
\unit \rightarrow \aa \, \nc^{2} \ \ {\rm as} \ \ \nc \to \infty
\hskip1.5cm {\rm and} \hskip1.5cm \unit \rightarrow -\bb \ \ {\rm as} \ \
\nc \to 0 \, .
\label{unite}
\end{equation}
It is worth noting that $\unit$ can be chosen to be a linear function
of $\nc^{2}$, $\, =\aa\nc^2-\bb$.  More generally, in what follows we
assume that $\unit$ is an analytic function of $\nc^2$.  This
assumption is not dictated by any fundamental principles but rather,
is appealing for the simplicity of its consequences.

As a first example, let us look at the non-Abelian running coupling
constant. It satisfies
\begin{equation} {1 \over \al}\,{d\alpha_s\over d \ln Q^2}=
- \beta_0 {\alpha_s\over \pi} - \beta_1 {\alpha_s^2\over \pi^2} +\cdots 
\end{equation} with ($T=1/2$, $\cf=(\nc^2-1)/2\nc$)
\begin{equation}
\beta_0 = {11 \over 12 }\nc\, -\, {1 \over 3}T \nf\, , \hskip 2cm
\beta_1 = {17 \over 24 }\nc^2\, -\, \left({5\over 12 } \nc +
{\cf\over 4 }\right) \,T \nf\, .\label{betafunct}
\end{equation}

Under the rescaling (\ref{rescaling}), these equations
become\footnote{For our specific choice of rescaling
$\unit=\unit_s=\unit_f$, ${\widehat T}= T \times (\unit_f/\unit_s)$
has the value $1$.}
\begin{equation}
 {1 \over \alh}\,{d\widehat \alpha_s\over d\ln Q^2} = - \widehat
\beta_0 {\widehat\alpha_s\over \pi} - \widehat \beta_1 {\widehat
\alpha_s^2\over \pi^2} + \cdots \qquad{\rm with} \qquad \widehat
\beta_n = {\beta_n\times\, \left({2\nc \over \unit}\right)^{n+1}}
. \label{runninghat}
\end{equation}

Explicit values for $\widehat \beta_0$ and $\widehat \beta_1$ are 
\begin{equation}\widehat \beta_0 = {11 \over 12 } \nch - {\nfh 
\over 3 }\, , \hskip 2cm {\widehat \beta}_1 = {17\over 24} \nch^2 -
\left({5 \over 12} \nch + {\cfh \over 4}\right) \nfh
\label{betahat}
\end{equation} 
where for conciseness, we have introduced $\nch = 
\nc \times 2\nc/\unit$ and $\cfh = \cf\times 2\nc/\unit$ which have
values $0$, $1/\bb$ at $\nc = 0$ and values $2/\aa$, $1/\aa$ at $\nc =\infty$
respectively.

The expressions (\ref{runninghat}) and (\ref{betahat}) are analytic 
functions of $\nc^2$ and are finite for all values of $\nc$ except 
possibly at zeroes of $\unit$.  A worth-noting peculiarity of these 
expressions is the change of sign of ${\widehat \beta}_0$ for 
sufficiently small values of $\nc$, conferring to the rescaled 
quantities an Abelian-like character in the vanishing-$\nc$ limit.  
These features are not accidental as we establish later.

In the particular limit $\nc\to 0$,
${\widehat \beta}_0$ and ${\widehat \beta}_1$ assume the values
\begin{equation}
{\widehat \beta}_0 = - {\nfh \over 3 }\, , \hskip 2cm 
\widehat \beta_1 = - \cfh \,{\nfh \over 4 } .
\end{equation}  
These values coincide with the values for QED with $\nfh$ fermions if
$\cfh$ takes its QED value, namely $\cfh=1$.\footnote{This and the
above statements also apply to the coefficient ${\widehat
\beta}_2$.}  This can be achieved for the specific choice
$\unit=\nc^{2}-1$ ($ \aa=\bb=1$).  Another curiosity about the
particular rescaling $\unit=\nc^{2}-1$, is that the Abelian character
turns on as $\nc$ goes across the ``free-fermion" value $\nc=1$, which
happens to be a zero of $\unit$ and at which value all rescaled
quantities are ambiguously defined.\footnote{Although they are
well-defined everywhere else in the complex $\nc^{2}$-plane.}

Next, we consider the ratio of the annihilation cross section to the 
point-like limit. The latter is, in the ${\overline {MS}}$ scheme,
 \begin{equation}
R_{e^+ e^-}(Q^2) = \nc \sum_I^{\nf} Q^2_I \lbrace 1 +
{\alpha_s(Q)\over \pi} F_2 +{\alpha_s^2(Q)\over
\pi^2} F_3 + \cdots \rbrace + \cdots 
\end{equation} 
where \(F_2 = {3\over 4} C_F\) and~\cite{ss}

\begin{equation}
F_3 = -{3\over 32} \cf^2 + \left({123 \over 32} - {11\over 4}
\zeta(3)\right) \nc\cf + \left( -{11\over 8} + \zeta(3)\right)\cf T
\nf.
\end{equation}

In terms of $\alh$ and $\nfh$, these equations become\footnote{The overall 
rescaling was chosen to factor-out an overall $\nc$-dependence.}
\begin{equation}
  R_{e^+ e^-}(Q^2)/\unit = 
 \sum_I^{\nfh} Q^2_I \lbrace 1 +{\alh(Q)\over \pi} \widehat F_2 
 +{\alh^2(Q)\over \pi^2}\widehat F_3 + \cdots \rbrace + \cdots  
 \end{equation}

with $\widehat F_n = F_n\times\,
\left(2\nc/\unit\right)^{n-1}$. Specifically,

\begin{equation} 
\widehat F_2 = {3\over 4}\cfh \ \ {\rm and} \ \ \ \widehat F_3 = -{3
\over 32}\cfh^2 + \left({123 \over 32} - {11\over 4} \zeta(3)\right)
\cfh\nch + \left( {-11\over 8} + \zeta(3)\right)\cfh\nfh\, .
\end{equation}

The coefficients $\widehat F_2$ and $\widehat F_3$ share the same
features as the coefficients of the $\beta$-functions:
They are analytic function in $\nc^{2}$ (and consequently in $\unit$),
and they equally suggest an Abelian character to the small-$\nc$
limit since they exactly reproduce the expected QED results for $\nfh$
flavors~\cite{ss} when $\cfh$ is set to $1$ and $\nch$ to $0$.  The
coefficients of $\al^3$ in the expansion above has been computed and
the corresponding $\widehat F_4$ also coincides with its QED
counterpart, but its expression is too lengthy to be reproduced
here~\cite{ss}.

The identification with QED holds in general as one introduces higher 
order Casimir (cubic, quartic, \ldots) in the expressions, providing 
that all rescaled Casimirs are formally replaced with their QED 
counterparts ($\nch\to0,\,\cfh\to1, \ldots$).  However, there is no 
unique choice of the value $\unit(\nc\to0)=-\bb$, which reproduces the 
QED limit for all rescaled colored objects.  To illustrate this point, 
let us consider the cubic Casimir $\dabc = (\nc^2-1)(\nc^2-4)/16\nc$ 
which arises at order $(\al^3, \nc^2)$.  It contributes to the 
$R$-ratio a term~\cite{ss}
\begin{equation}
 R_{e^+ e^-}(Q^2) = \ldots + {\al^3(Q)\over\pi^3}\left(\sum_I^{\nf} 
 Q_I\right)^{2}
\,\dabcf\, \left( {11\over 12} - 2 \zeta(3)\right).
\label{Rratiodabc}
\end{equation}
The corresponding contribution to the
rescaled ratio is
\begin{eqnarray}
 R_{e^+ e^-}(Q^2)/\unit &=& \ldots + 
{\alh^3(Q)\over\pi^3}\left(\sum_I^{\nfh} Q_I\right)^{2} \,\dabcfh\, \left( 
{11\over 12} - 2 \zeta(3)\right)  \label{Rhatratiodabc} \\
{\rm with}\qquad\dabcfh
 &=& \left({1 \over 2}{\nc^2-4 \over\unit}\,{ \nc^{2}-1 \over 
 \unit}\right)\, .
\nonumber
\end{eqnarray}
The QED value of the R-ratio is given by Eq. (\ref{Rhatratiodabc}) 
with the substitution $\dabch \to 1$, and is $\bb^2/2$ times the 
value of the $\nc\to 0$ limit of Eq. (\ref{Rhatratiodabc}).  This 
expected departure from QED is calculable from the topology of the 
diagrams as we demonstrate below (See also Appendix B).

We have now enough examples to profile the general color structure of
a physical quantity under the rescaling proposed in Eq.
(\ref{rescaling}).  The color structure is an analytic function of
$\nc^2$ or, after solving $\nc^2$ as a function of $\unit$, it is an
analytic function of $\unit$, with finite values at $\nc=0$ and
$\infty$.  Furthermore, for sufficiently small $\nc$, it has an
Abelian character, in particular, it formally coincides with its QED
value upon substitution of irreducible group invariants with their QED
counterparts ($\nch\to0,\,\cfh\to 1, \dabch\to 1, \ldots$), a
feature which may provide a powerful test of QCD calculations with
$\nf$ flavors by comparing them with their corresponding QED results
with $\nfh$ flavors.  This identification can be achieved with a
particular choice of rescaling $\unit(\nc\to0)=-\bb=-1$, to an order
in $\al,\,\nf$, which depends on the quantity under consideration.  At
higher order, there is a systematic deviation which reflects the
topology of the diagrams involved (See below).

Another feature somewhat hidden in the examples above is a restriction
on the possible dependence on $\unit$ at a given order of $\al,\,\nf$.
To formulate this feature, it is convenient to associate a
``dimension" $q$ to a color group theoretic object $\col{}{}$, which
is essentially a count of the power of $\nc$ in the large $\nc$ limit:
$\col{}{} \sim \nc^{q}, \ \nc\to \infty$.  The choice of rescaling of
$\al$ and $\nf$ made in Eq. (\ref{rescaling}) compels us to
associate a dimension of $-1$ to $\al$ and $+1$ to $\nf$ (\emph{ hat}
quantities are dimensionless).  The statement goes as follows:

\parbox{13cm}{\it \noindent A quantity at order $\al^{n}$ and
$\nf^{\lf}$ can only involve a product of group theoretic objects
whose ``dimension" is less or equal to $2+n-\lf$; The
equality is achieved by, and only by, planar diagrams.  Namely,}

\begin{equation}
\col{n,\lf}{} = \cdots + \#_{(4d)}\ \al^n \nf^{\lf}\, \left(
\nc^\alpha \cf^\beta \left[\dabcf\right]^\gamma \cdots\right)
\ + \ldots \,  .
\label{dimensionality}
\end{equation}
\begin{equation}
\mbox{with} \qquad \left\{ { \ \ \alpha+\beta+3\gamma + \cdots <
 {2+n-\lf \qquad \mbox{\it nonplanar diagrams}} \atop
 \!\!\!\!\!\!\!\!\alpha+\beta+3\gamma + \cdots = {2+n-\lf \qquad
 \mbox{\it planar diagrams}} }\right. \label{dimequation}
\end{equation} 

The $\cdots$ in Eq. (\ref{dimequation}) represent the
contribution of higher order Casimirs.  It is worth emphasizing that
this statement, not surprising in the large-$\nc$ limit, is a
nontrivial statement valid for arbitrary values of $\nc$, in
particular, $\#_{(4d)}$ contains the spacetime dependence without any
reference to color degrees of freedom.  This ``homogeneity'' rule may
constitute yet another tool to test the consistency of QCD
calculations and allows one to pinpoint easily the contributions from
planar diagrams. The examples previously given simply illustrate the
nature of the theorem as they only involve the computation of planar
diagrams (the ``dimensions" of $\nc,\, \cf$ and $\dabc$ being
$1,\,1$ and $3$, respectively). Thus, the ``dimension'' of
$\col{n,\lf}{}$ including at least one planar diagram is $q=2$.

The small-$\nc$ limit is obtained from Eq. (\ref{dimensionality}) 
after proper rescaling of $\al$ and $\nf$.  That the rescaled 
expression does not diverge as powers of $1/\nc$ in the limit 
$\nc\to0$, guarantees the finiteness of the vanishing-$\nc$ limit and 
is proven in the next section.

\subsection*{General Analysis}

To understand better the ``homogeneity'' rule for all $\nc$ and the
Abelian-ity of the vanishing limit $\nc=0$, as well as other features
illustrated by the above examples, we now consider the diagrammatics
from another perspective.  It is convenient to use the following
representation of the $\sun$ algebra
\begin{equation} \sup{T}{\alpha\,i}{\beta\,j}\,=\, {1\over 2}
\,\left(\delta^{\alpha}_j \delta^i_{\beta} \,-\,{1 \over
\nc}\,\delta^{\alpha}_{\beta}\delta^i_j\right)\,.
\label{Tgenerators} 
 \end{equation} 
The components $\delta^{\alpha}_j \delta^i_{\beta}$ of the generators 
$\sup{T} \alpha\beta$ are the $\nc^2$ generators of $U(\nc)$ and 
constitute the ``double lines" of the large-$\nc$ limit.  The second 
components $-\,{1 \over \nc}\,\delta^{\alpha}_{\beta}\delta^i_j$ are 
required to ensure tracelessness of $\sup{T} \alpha\beta$ and, as they 
have a trivial color structure, we refer to them in a diagrammatic 
context as to ``photon lines"; they play an important role in the 
small-$\nc$ limit.

A given perturbative physical quantity $\mathcal R$ derives its color
dependence from Feynman diagrams with no external legs, $\rl$,
involving $n$ exchanged gluons and $\lf$ fermion loops (``bubble"
diagrams). After all gluons are replaced according to the rule
(\ref{Tgenerators}), a ``bubble" diagram $\rl$ is equivalent to the sum
of $2^{n}$ \emph{component-diagrams} $\dol{n,\lf}{e_i}$, labelled with
an index $i$, involving $n-e_i$ ``photon lines" and $e_i$ ``double
lines": $\rl=\sum_{i}^{2^{n}}\dol{n,\lf}{e_i}$ (Fig.  $1$). One such
component-diagram $\dol{n,\lf}{e_i}$ is said to be \emph{connected} if
all of its parts are connected with ``double lines", otherwise, it is
said to be \emph{disconnected} and is composed of $d_i$ connected
graphs $\col{\lf_{i}^\alpha}{e_i^\alpha}$, each composed of
$\lf^\alpha_{i}$ fermion loops and $e_i^\alpha$ ``double lines":
$\dol{n,\lf}{e_i}=\prod_{\alpha}^{d_i}\col{\lf^\alpha_i}{e_i^\alpha}$
(Fig.  $1$).  These $d_i$ connected graphs
$\col{\lf^\alpha_i}{e_i^\alpha}$ are implicitly attached to each other
with ``photon lines"; furthermore, each of them has a topology of a
surface~\cite{thooft}, characterized by $H_i^\alpha$ handles,
$B_i^\alpha$ holes and an Euler number $\chi_i^\alpha =
2-2H_i^\alpha-B_i^\alpha$.  To a component-diagram $\dol{n,\lf}{e_i}$,
we associate a total Euler number $\chi_i$ which is the sum of the
Euler numbers of its connected graphs, $\chi_i= \sum_\alpha
\chi_i^\alpha$.

\vspace{.5cm}
\begin{figure}[htb]
\begin{center}
\leavevmode
{\epsfxsize=4.5in\epsfbox{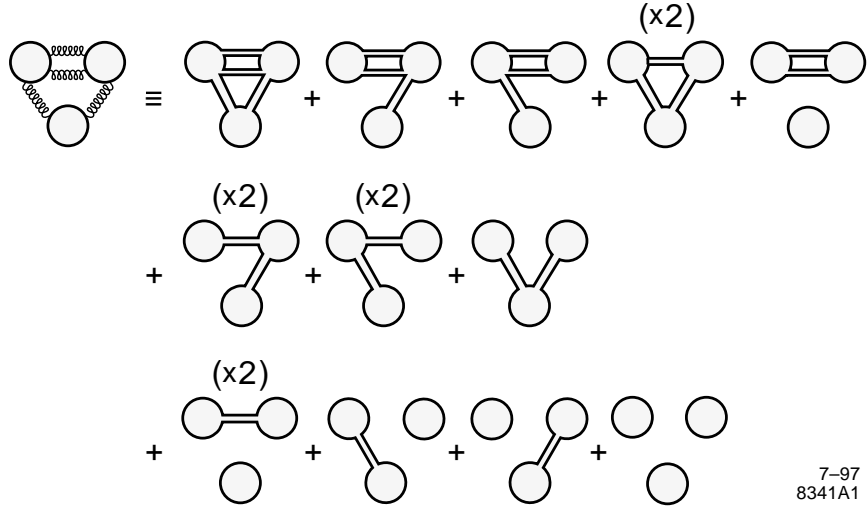}}
\end{center}
\caption[*]{\small
This example illustrates the terminology introduced in the text.
This particular ``bubble'' diagram has $n=4$, $\lf=3$ and
decomposes into $2^4$ component-diagrams: $(1\times) \, \dol{4,3}{4},
\, (4\times) \,\dol{4,3}{3}, \, (6\times) \,\dol{4,3}{2}, \, (4\times)
\,\dol{4,3}{1},\, (1\times) \,\dol{4,3}{0}$. Some of these
component-diagrams are disconnected; this is the case for the last
component-diagram of the first row which decomposes into two connected
graphs with $2$ loops and $1$ loop, respectively: $ \dol{4,3}{2}\,=
\col{2}{2} \times \col{1}{0}$.  The index $\index$ of these
component-diagrams are $2,1,1,1,1$ and $1$ for the six contributions of
the first row and is $0$ for the remaining $10$ contributions occupying
the last three rows. The large-$\nc$ limit is controlled by the first
diagram with maximal $\index=2$, while the vanishing-$\nc$ limit is
dominated by the ``tree''-graphs characterized by $\index = 0$. }
\label{fig1} 
\end{figure}

Having introduced this terminology, we now express the 
$\nc$-dependence of a physical quantity.  The color dependence 
$\col{n,\lf}{}$ of a ``bubble'' diagram $\rl$ is, expressed in terms 
of $\alh$ and $\nfh$, 

\begin{eqnarray}
 \col{n,\lf}{} &=& \, \alh^{n} \, \nfh^{\lf}
\ \unit^{\lf-n} \sum_{i}^{2^n} (-1)^{n-e_i}
\left(\nc^{2}\right)^{\index_i} \nonumber \\ &=&\,
\alh^{n} \, \nfh^{\lf} \ \unit^{\lf-n} \sum_{i}^{2^n}
(-1)^{n-e_i}\ \prod_{\alpha=1}^{d_i} \
\left(\nc^{2}\right)^{\index_i^\alpha}
\label{generalunit}
\end{eqnarray}
where $\unit$ is viewed as an analytic function of $\nc^2$, and we have
introduced a topological index $\index_i^\alpha$ defined as 

\begin{equation}
\index_i^\alpha = e_i^\alpha+(\chi_i^\alpha-\lf^\alpha_i)/ 2 \, .
\label{xalphadefined}
\end{equation} 
The total index $\index_i$ for a component-diagram is 
\begin{equation}
\index_i = \sum_\alpha^{d_i} \index_i^\alpha =  e_i+(\chi_i-\lf)/ 2 \, .
\label{xdefined}
\end{equation} 

Using Eqs. (\ref{generalunit}-\ref{xdefined}) above, we can
readily demonstrate all results we inferred from the earlier examples,
namely, $(1)$ analyticity in $\nc^2$ of $\col{n,\lf}{} $, $\, (2)$
``homogeneity'' as formulated in Eq. (\ref{dimensionality}), and
$(3)$ Abelian behavior in the vanishing $\nc\to0$ limit.

To establish the analyticity of $\col{n,\lf}{}$ as a function of
$\nc^2$ (or $\unit$), we need only to observe that
$(\chi_i^\alpha+\lf^\alpha_i)/2$ is an integer, $\leq 1$, the equality
being satisfied when the corresponding connected graph
$\col{\lf^\alpha_i}{e_i^\alpha}$ has a planar topology.

In order to establish the two other points, we observe that the index
$\index_i$ defined in (\ref{xdefined}) for each component-diagram
$\dol{n,\lf}{e_i}$ in Eq. (\ref{generalunit}) is a nonnegative
integer (cf. Appendix A). In particular,
\begin{enumerate}
\item[(1)]

$\index_{i}$ is maximal and equal to $ n +(\chi_i-\lf)/ 2$ when the 
component-diagram $\dol{n,\lf}{e_i}$ is itself connected ($d_i=1$) with 
$e_i^\alpha = n$, corresponding to replacing all gluons in the 
``bubble'' diagram with ``double lines" (See Fig. $1$).
\item[(2)]

$\index_{i}$ is minimal and zero when each corresponding
exponent $\index_i^\alpha$ is zero.  This case arises when each
of the connected graph $\col{\lf^\alpha_i}{e_i^\alpha}$ has its
$\lf^\alpha_i$ fermion loops interconnected with ``double lines" such
that \emph{there is no closed path}; in particular, there is no more
than one ``double line" between two fermion loops (See Fig.  $1$).  An
alternate formulation amounts to saying that every corresponding
connected graph has a \emph{``tree" shape}.  The ``tree"-graph has a
planar topology characterized by one face ($F=1$), no handle ($H=0$)
and the number of holes equal to the number of fermion loops
($B=\lf$).
\end{enumerate}

We can now proceed further.  In order to demonstrate the
``homogeneity'' rule stated in Eq. (\ref{dimensionality}), we
consider the large-$\nc$ behavior of Eq. (\ref{generalunit}).

The large-$\nc$ behavior is readily seen to arise from the
component-diagram with maximal index $\index_i$, that is, from the
connected component-diagram composed of only one connected graph
obtained from replacing all gluons with ``double lines" as explained
in item $(1)$.  In that case, $\col{n,\lf}{}$ is a Laurent series in
$\nc^{2}$ whose leading term is
\begin{equation}
\left.\col{n,\lf}{}\right|_{\nc\to\infty} = \, \alh^{n} \, \nfh^{\lf} 
\  \aa^{\lf-n} \left(\nc^{2}\right)^{(\chi_{\rm max}+\lf)/ 2}\, .
\label{generalinfinity}
\end{equation}
As noted before, the exponent $(\chi_{\rm max}+\lf)/ 2$ is $\leq 1$, 
the equality being achieved for a ``bubble'' diagram with planar 
topology, in which case, the physical quantity scales as $\nc^2 \sim 
\unit$.  This is expected in the context of a membrane interpretation of 
the color space in the large $\nc$ limit, for which physical 
quantities naturally scale as the area of the membrane, $\propto 
\nc^2$~\cite{hoppetal}.  This fact is also at the origin of the 
``scaling'' law stated in Eq. (\ref{dimensionality}, 
\ref{dimequation}) which is the statement that $\alpha,\beta, 
\gamma,\ldots$ defined in Eq. (\ref{dimensionality}), satisfy
\begin{equation}
\alpha+\beta+3\gamma+\cdots\, =\, \chi_{\rm max} + \lf \, , 
\label{dimequationChi}
\end{equation}
a number fixed by the number of loops $\lf$ and the topology of the
``bubble'' diagram.  The inequality in Eq. (\ref{dimequation})
translates into the known inequality $\chi+\lf \leq 2$, which
saturates only for planar diagrams.

In the vanishing limit ($\nc\to0$), the component-diagrams that
dominate the color factor are those which are the most suppressed in
the large-$\nc$ limit.  More precisely, they are those for which the
index $\index_i$ is minimal and vanishes.  According to item $(2)$ above,
this happens for those component-diagrams composed of tree-graphs.
The value of the color factor of a ``bubble'' diagram becomes a
polynomial in $\nc^{2}$, whose nonvanishing term is
\begin{eqnarray}
\left.\col{n,\lf}{}\right|_{\nc\to0} &=& \, \alh^{n} \, \nfh^{\lf} 
\times { {\nc\to0} \choose {\rm QED} } \nonumber \\ \mbox{with}\qquad 
{ {\nc\to0} \choose {\rm QED} } &=& \,\bb^{\lf-n} 
\sum_i^{\rm tree \atop \rm diagrams }\left(-1\right)^{\lf-e_i}\,. 
\label{tozero} 
\end{eqnarray}
The factor ${ {\nc\to0} \choose {\rm QED} }$ is the ``offset'' with
respect to the QED value.  Setting ${ {\nc\to0} \choose {\rm QED} }$
to $1$ formally in Eq. (\ref{tozero}) amounts to setting
$\nch\to 0$ and $\cfh, \dabcfh, \ldots \, \to \, 1$ in the examples
shown earlier and formally reproduces the corresponding QED
calculation with $\nfh$ flavors.  However, this is not an analytic
procedure, since there is no choice of value for $\bb$ which sets the
``offset'' ${ {\nc\to0} \choose {\rm QED} }$ to $1$ for all diagrams.

The factor ${ {\nc\to0} \choose {\rm QED} }$ is very easy to compute
for an arbitrary diagram.  The calculation is a simple combinatoric
exercise which does not make use of the full structure of the original
diagram because the sum involved is limited to tree-graphs; we provide
some examples in Appendix B. In particular, the non-Abelian
structure of the original diagram such as gluon loops and cubic- and
quartic-gluon vertices are all suppressed by powers of $\nc^{2}$ in
the vanishing limit.  Furthermore, because the endpoints of a ``double
line" involved in the ``tree"-graphs are each attached to one face,
the corresponding generators belong to the Cartan sub-algebra of
$\sun$.  Hence, all remnant  of non-Abelianity has vanished in the
$\nc\to0$ limit.  More precisely,
\begin{equation}
\lim_{\nc\to0}\,\sun \ = \ \lim_{\nc\to0}\,\left[\u1\right]^{\nc-1}\, .
\label{abelianlimit}
\end{equation}
This result confirms the change of sign of the first coefficient
${\widehat \beta}_0$ of the beta-function for sufficiently small
values of $\nc$.  The color factor of a ``tree"-graph scales as
$\nc^{0}$, i.e., it has ``dimension'' 0.  The zero-``dimension'' of a
``tree"-graph suggests a ``point-like'' structure which reflects its
topological properties, namely $\lf$ punctures on a surface (the
number of holes $B = \lf$) interconnected so to form a mesh of zero
area (the mesh has no closed loops so that its number of faces is
minimal $F=1$).  These properties are to be contrasted with the know
non-Abelian limit
\begin{equation}
\lim_{\nc\to\infty}\,\sun \ = \ \lim_{\nc\to\infty}\,\un\, .
\label{thooftlimit}
\end{equation}
In this limit, the color structure is controlled by planar diagrams of
``dimension" $2$~\cite{thooft, veneziano}, and the natural objects are
two-dimensional~\cite{hoppetal}.

A question which comes to mind is what is the dynamics underlying the
$\nc\to0$ limit of an $\sun$ gauge theory, represented by the
expansion (\ref{tozero}).  To provide a full answer to this question
may require the introduction of a prescription to construct the
$\nc\to0$ limit of a Feynman diagram with external legs, a
mathematical challenge that we haven't addressed in the present work.
One may naively anticipate, however, that the $\nc\to0$ limit is not a
quantum field theory defined in the usual sense: The $\nc(\nc-1)$
off-diagonal gluons disappear from the theory, degrees of freedom
which might otherwise be needed to insure unitarity and
positivity.\footnote{Speculatively, the quantum theory of an open
system might be a more appropriate formulation of this limit as also
suggested by its relevance to condensed matter systems.} The
interpretation of the dynamics of the Abelian $\nc\to0$ limit is more
likely to take a more geometrical form in light of the point-like
nature of this limit on a surface in color space.  Along these lines,
while the large-$\nc$ limit relates the dynamics of certain $\sun$
theories to the tension of a membrane~\cite{hoppetal}, the small-$\nc$
limit plausibly relates its dynamics to interlinked punctures on a
surface.\footnote{It is tempting to interpret them as a cluster of
point-like objects.} Because these points (fermion loops) are connected
pairwise so as not to form a closed path which would otherwise be
suppressed by $\nc^{2}$, the dynamics of this limit evidently does not
depend on the tension of the surface.

It is intriguing to observe from Eq. (14) that the small-$\nc$ limit
receives contributions from higher order Casimirs of $\sun$ at
arbitrary $\nc$, the number of which increases as $\nc$ increases. For
example, the cubic Casimir $\dabc$ which arises for $\sun$ with integer
$\nc \geq 3$ and which vanishes for $\nc=$ 1 and 2, does contribute to the
$\nc\to0$ limit.  This relation of the small-$\nc$ limit to the
algebraic structure of the large-$\nc$ limit is an important
consequence of the extension of the rescaled group-theoretic objects of
$\sun$ viewed as an analytic function of $\nc^2$.  This may be
indicative of a more profound connection between the large- and
small-$N_c$ theories.

It is remarkable that a topological expansion determines the color 
structure of $\sun$ in both the large- and the small-$\nc$ limits.  
This observation and the analyticity properties demonstrated above are 
in our view in strong support of a yet to be uncovered deeper and more 
universal structure underlying $\sun$ gauge theories for arbitrary 
values of $\nc$.

\subsection*{Acknowledgments} We acknowledge Eric Sather for making 
significant contributions at an early stage of this work, Michael E.
Peskin and Lance Dixon for their insightful remarks, and Michael Dine
and Michael Melles for making useful comments on an early manuscript.

\newpage

\subsection*{Appendix A} In this appendix, we sketch the proof of the 
stated properties of the index $\index_i^\alpha = 
{e_i^\alpha+(\chi_i^\alpha-\lf^\alpha_i)/ 2}$ defined in 
(\ref{xalphadefined}) which is characteristic of the topology of a 
connected graph $\col{\lf^\alpha_i}{e_i^\alpha}$ composed of 
$\lf^\alpha_i$ fermion loops and $e_i^\alpha$ ``double 
lines".

All stated properties of the index $\index$ can be derived from the
following rules:
\begin{itemize}

\item[(1)] If two connected graphs $\col{1}{}$, $\col{2}{}$ are
interconnected with one ``double line", the newly formed connected
graph $\col{12}{}$ satisfies $\index_{12}=\index_1+\index_2$.

\item[(2)] If one ``double line" is attached to a connected graph
$\col{1}{}$, the new connected graph $\col{}{}$ has an index $\index =
\index_1+1$.

\item[(3)] The simplest connected graph composed of one fermion loop
and no  ``double line" has $\index=0$.

\end{itemize}

The proof of the above statements rely on the relation $\chi = F-E+V$,
where $F$ is the number of faces, $E$, the number of edges and $V$ the
number of vertices. The first item above relies on the fact that when
adding one ``double line", $F \to F-1$, $E \to E+3$ and $V \to V+2$;
similarly for the second item with $F \to F+1$ instead; the last
item is a consequence of a fermion loop being characterized by $F=1$
and $E=V=0$.

In order to demonstrate that $\index \geq 0$, it suffices to observe that 
all graphs, connected or not, can be constructed from more elementary 
ones according to the rules above.

\newpage

\subsection*{Appendix B}
In this appendix we illustrate with a few examples the simplicity of
the sum over ``tree"-graphs in Eq. (\ref{tozero}):
\begin{eqnarray}
\left.\colh{n,\lf}{}\right|_{\nc\to0} &=& \, \alh^{n} \, \nfh^{\lf} 
\times { {\nc\to0} \choose {\rm QED} }  \nonumber\\ \mbox{with}\qquad 
{ {\nc\to0} \choose {\rm QED} } &=& \,\bb^{\lf-n} \times
\sum_i^{\rm tree \atop \rm diagrams }\left(-1\right)^{\lf-e_i}\, . 
\nonumber \\ \nonumber
\end{eqnarray}

Recall that all diagrams with non-Abelian couplings contribute to zero
in the small-$\nc$ limit.  The first nontrivial example is a
``bubble'' diagram with one fermion loop and an arbitrary number of
gluons.  The only ``tree"-graph is the fermion loop itself with
$e_i=0$; hence, ${ {\nc\to0} \choose {\rm QED} }$ becomes

\begin{eqnarray}
{ {\nc\to0} \choose {\rm QED} } &=& \,\bb^{1-n} \times
\sum_i^{\rm tree \atop \rm diagrams }\left(-1\right)^{1-e_i} \nonumber
\\ 
&=& \, \bb^{1-n} \times \left[ \ \  \onegraph \ \  \right]
 \nonumber \\
&=& \, \bb^{1-n} \times \left( \   -1  \  \right) \, .
\label{oneloop}
\end{eqnarray}

As a second example, consider the most general two-fermion loop diagram 
with $n$ gluons, $p$ of them interconnecting the two loops.  For that 
case, there are two types of ``tree"-graphs and one obtains

\begin{eqnarray}
{ {\nc\to0} \choose {\rm QED} } &=& \,\bb^{2-n} \times \sum_i^{\rm
tree \atop \rm diagrams }\left(-1\right)^{2-e_i} \nonumber \\
 &=& \, \bb^{2-n} \times \left[ \ \onegraph \ \onegraph \ \ + \ p \
\twograph \ \right] \nonumber \\ &=& \, \bb^{2-n} \times \left( \ 1 \
- \ p \ \right) \, .
\label{twoloop}
\end{eqnarray}

Now we consider the most general three-fermion loops diagram with $n$
gluons, each pair of fermion loops being interconnected with $p,q$ and
$r$ gluons respectively.  For that case there are three topologically
distinct types of ``tree"-graphs and

\begin{eqnarray} 
{ {\nc\to0} \choose {\rm QED} } &=& \,\bb^{3-n} \times \sum_i^{\rm
tree \atop \rm diagrams }\left(-1\right)^{3-e_i} \nonumber \\ &=& \,
\bb^{3-n} \times \left[ \ \onegraph \ \ \onegraph \ \ \onegraph \right.
\nonumber \\ &\qquad& \left. + \
(p+q+r) \ \twograph \ \onegraph \ + \ (pq+qr+rp) \ \threegraph \
\right] \nonumber \\ &=& \, \bb^{3-n} \times \left( \ -1 \ + \
(p+q+r)\ - (pq+qr+rp) \ \right) \nonumber \\ &=& \,\bb^{3-n} \times
\left( \ (p-1)(q-1)(r-1) \ - \ pqr \ \right) \, .
\label{threeloop}
\end{eqnarray}

This example easily generalizes to $n$ gluons and $\lf$ fermion loops
interconnected so to form a ring, i.e., interconnected with
$p_{(12)},\ldots , p_{(\lf 1)} $ gluons:
\begin{equation} 
{ {\nc\to0} \choose {\rm QED} } \, = \,\bb^{\lf-n} \times \left( \
(p_{(12)}-1) \cdots ( p_{(\lf 1)}-1) \ - \ p_{(12)}\cdots p_{(\lf 1)}
\ \right) \, .
\label{ring}
\end{equation}

Finally, the most general ``bubble'' diagram with $\lf$ fermion 
loops labeled with $i= 1, \ldots, \lf$ and $n$ gluons distributed 
such that $p_{(ij)}$ of them interconnect the fermion loops $i$ and $j$, 
will have a ``offset'' factor ${ {\nc\to0} \choose {\rm QED} }$  given by 
the sum
\begin{eqnarray} 
{ {\nc\to0} \choose {\rm QED} } &=& \,\bb^{\lf-n} \times \sum_i^{\rm
tree \atop \rm diagrams }\left(-1\right)^{\lf-e_i} \label{manyloop} \\ &=&
\,\bb^{\lf-n} (-1)^{\lf} \,\times \left[ \ \onegraph \ \cdots \onegraph \ +
\ (p_{(12)}+\cdots+p_{(ij)}+\cdots) \ \twograph \ \onegraph \ \cdots
\onegraph \right.  \nonumber \\ &+& \ \left.  (p_{(12)}p_{(34)}+
\cdots + p_{(ij)}p_{(kl)}+ \cdots) \ \twograph \ \twograph \ \onegraph
\ \cdots \ \onegraph \ + \ \cdots \right] \nonumber \\ &=& \,
\bb^{\lf-n} (-1)^{\lf} \, \times \left( \ 1 \ - \ \sum_{(ij)}
p_{(ij)}\,+\, \sum_{(ij)\neq(kl)} p_{(ij)}p_{(kl)}\, - \, \cdots \
\right) \, .  \nonumber
\end{eqnarray}
Using Eq. (\ref{manyloop}) above, one can show that
\begin{equation}
\sum_i^{\rm tree \atop \rm diagrams }\left(-1\right)^{\lf-e_i} \,\leq \, 
0 \, .
\label{signofsum}
\end{equation}
With the help of Eq. (\ref {signofsum}) one can infer that ${
{\nc\to0} \choose {\rm QED} } \leq 0 $ when $\unit(\nc\to0) = - \bb <
0$.

\end{document}